\documentclass[a4paper,12pt]{article}

\usepackage{amsfonts}
\usepackage{amsmath}
\usepackage{amssymb}
\usepackage{amsthm}
\usepackage{mathtools}
\usepackage{cite}
\usepackage{xspace}
\usepackage{float}
\usepackage{subcaption}
\usepackage{color}
\usepackage{tikz}
\usepackage[inline]{enumitem}
\setlist{topsep=2mm,itemsep=1.5mm}
\usepackage{booktabs}
\usepackage{multirow}
\usepackage{hyperref}
\usepackage[margin=30mm]{geometry}
\usetikzlibrary{shapes,positioning,patterns}
\tikzset{%
  detail/.style={%
    draw, rectangle split, rectangle split parts=#1,
    rectangle split draw splits=true
  },
  nodeid/.style={%
    circle, draw, minimum size=4.5ex, very thick, inner sep=2pt
  },
  edgeid/.style={%
    nodeid, draw=white, fill=white
  },
  relationship/.style={%
    draw, very thick, -latex
  }
}
\makeatletter
\pgfdeclarepatternformonly[\GridSize]{MyGrid}{\pgfqpoint{-1pt}{-1pt}}{\pgfqpoint{4pt}{4pt}}{\pgfqpoint{\GridSize}{\GridSize}}%
{
  \pgfsetcolor{\tikz@pattern@color}
  \pgfsetlinewidth{0.3pt}
  \pgfpathmoveto{\pgfqpoint{0pt}{0pt}}
  \pgfpathlineto{\pgfqpoint{0pt}{3.1pt}}
  \pgfpathmoveto{\pgfqpoint{0pt}{0pt}}
  \pgfpathlineto{\pgfqpoint{3.1pt}{0pt}}
  \pgfusepath{stroke}
}
\pgfdeclarepatternformonly[\LineSep]{hLines}{\pgfpointorigin}{\pgfqpoint{100pt}{1pt}}{\pgfqpoint{100pt}{\LineSep}}%
{
  \pgfsetcolor{\tikz@pattern@color}
  \pgfsetlinewidth{0.3pt}
  \pgfpathmoveto{\pgfqpoint{0pt}{0.5pt}}
  \pgfpathlineto{\pgfqpoint{100pt}{0.5pt}}
  \pgfusepath{stroke}
}

\pgfdeclarepatternformonly[\LineSep]{vLines}{\pgfpointorigin}{\pgfqpoint{1pt}{100pt}}{\pgfqpoint{\LineSep}{100pt}}%
{
  \pgfsetcolor{\tikz@pattern@color}
  \pgfsetlinewidth{0.3pt}
  \pgfpathmoveto{\pgfqpoint{0.5pt}{0pt}}
  \pgfpathlineto{\pgfqpoint{0.5pt}{100pt}}
  \pgfusepath{stroke}
}
\makeatother

\newdimen\GridSize
\newdimen\LineSep
\tikzset{%
  GridSize/.code={\GridSize=#1},
  GridSize=3pt,
  LineSep/.code={\LineSep=#1},
  LineSep=3pt
}
\usepackage{relsize}
\usepackage{authblk}

\title{Formal Semantics of the Language Cypher \\ {\large Version 1.1 : core read-only fragment}}

\author[1]{Nadime~Francis\thanks{Affiliated with the School of Informatics at the University of
  Edinburgh during the time of contributing to this work.}}
\affil{Universit{\'e} Paris-Est, France}
\author[2]{Alastair~Green}
\affil{Neo4j}
\author[3]{Paolo~Guagliardo}
\affil{University of Edinburgh}
\author[3]{Leonid~Libkin}
\author[2]{Tobias~Lindaaker}
\author[3]{Victor~Marsault}
\author[2]{Stefan~Plantikow}
\author[2]{Mats~Rydberg}
\author[3]{Martin~Schuster}
\author[2]{Petra~Selmer}
\author[2]{Andr{\'e}s~Taylor}

\date{}
\usepackage{textcomp}
\usepackage[final]{listings}
\definecolor{darkblue}{rgb}{0.0, 0.0, 0.55}
\definecolor{darkred}{rgb}{0.8, 0.0, 0.0}

\lstdefinestyle{cypher}{%
  language=SQL,
  morekeywords={QUERY,MATCH,RETURN,WITH,OPTIONAL,GRAPH,OF},
  deletekeywords={YEAR,SIZE}
}
\lstdefinestyle{onedigit}{%
  numbers=left,
  numbersep=6pt,
  xleftmargin=\dimexpr\parindent
}
\lstdefinestyle{twodigits}{%
  numbers=left,
  numbersep=6pt,
  xleftmargin=\dimexpr\parindent+3.5pt
}
\lstnewenvironment{cypher}[1][]{%
  \lstset{#1,style=cypher}%
}{}

\newcommand{\sqlkw}[1]{\text{\ttfamily\bfseries\color{darkblue}#1}}
\newcommand\set[1]{\{#1\}}

\newcommand\sem[1]{{[\![ #1 ]\!]}}
\newcommand\code[2][]{\rootcode[#1]{#2}}

\newcommand\cypherlist[1]{\textsf{list}(#1)}
\newcommand\cyphermap[1]{\textsf{map}(#1)}
\newcommand\cypherpath[1]{\textsf{path}(#1)}
\newcommand\symbollist[1]{[#1]}
\newcommand\symbolmap[1]{\{#1\}}
\newcommand\mymid{\ {\big|}\ }
\newcommand{\rootcode}[2][]{\text{\normalfont#1\texttt{#2}}}
\newcommand{\kwcode}[1]{\rootcode[\color{darkblue}\bfseries]{#1}}

\newcommand\V{\mathcal{V}}

\newcommand\K{\mathcal{K}}
\newcommand\N{\mathcal{N}}
\newcommand\R{\mathcal{R}}
\newcommand\Lab{\mathcal{L}}
\newcommand\T{\mathcal{T}}
\newcommand\A{\mathcal{A}}
\newcommand\F{\mathcal{F}}

\newcommand\nv{\sqlkw{null}}
\newcommand\true{\sqlkw{true}}
\newcommand\false{\sqlkw{false}}

\newcommand\Z{\mathbb{Z}}
\newcommand\Nat{\mathbb{N}}
\newcommand\distinct{\varepsilon}
\newcommand\rigid{{\normalfont\textsf{rigid}}}
\makeatletter
\newcommand{\free}{\text{free}\@ifstar{\text{*}}{}}
\makeatother
\newcommand{\freevars}{\textsf{free}}

\newcommand{\OMIT}[1]{}

\newcommand\phantommid{\phantom{\mymid}}
\makeatletter
\newcommand{\grammartoken}[1]{\textsf{#1}}
\newcommand{\expr}{\grammartoken{expr}}
\newcommand{\retarg}{\@ifstar{\grammartoken{ret}^{\circ}}{\grammartoken{ret}}}
\newcommand{\query}{\@ifstar{\grammartoken{query}^{\circ}}{\grammartoken{query}}}
\newcommand{\clause}{\grammartoken{clause}}
\newcommand{\aggexpr}{\grammartoken{aggexpr}}

\newcommand{\name}{a}
\newcommand{\key}{k}
\newcommand{\lab}{\ell}
\newcommand{\props}{\grammartoken{map}}
\newcommand{\type}{t}
\newcommand{\iter}{\grammartoken{len}}
\newcommand{\qm}{\grammartoken{?}}
\newcommand{\keyexprtuple}{\grammartoken{prop\_list}}
\newcommand{\exprlist}{\@ifstar{\grammartoken{expr\_list}^\circ}{\grammartoken{expr\_list}}}
\newcommand{\labellist}{\grammartoken{label\_list}}
\newcommand{\typelist}{\@ifstar{\grammartoken{type\_list}^\circ}{\grammartoken{type\_list}}}
\newcommand{\pathpattern}{\@ifstar{\grammartoken{pattern}^{\circ}}{\grammartoken{pattern}}}
\newcommand{\nodepattern}{\grammartoken{node\_pattern}}
\newcommand{\relpattern}{\grammartoken{rel\_pattern}}
\newcommand{\pathpatternlist}{\grammartoken{pattern\_tuple}}
\makeatother

\newcommand{\dom}{\textsf{dom}}

\renewcommand{\emptyset}{\varnothing}

\newcommand{\nil}{{\normalfont\textsf{nil}}}
\newcommand{\ltor}{\ensuremath{{\rightarrow}}}
\newcommand{\rtol}{\ensuremath{{\leftarrow}}}
\newcommand{\undir}{\ensuremath{{\leftrightarrow}}}
\newcommand{\getk}[2]{\ensuremath{\iota(#1,#2)}}
\newcommand{\src}{\textsf{src}}
\newcommand{\tgt}{\textsf{tgt}}
\newcommand{\match}{\textsf{match}}

\makeatletter
\newcommand{\vm@date@separator}{\hspace*{0.15ex}\rule[0.4\vm@date@height]{1ex}{0.07\vm@date@height}\hspace*{0.15ex}}
\newcommand{\vmdatefont}[1]{#1}
\newcommand{\isotoday}{%
  \vmdatefont{
    \newdimen\vm@date@height%
    \setbox0=\hbox{0123456789}%
    \vm@date@height=\ht0 \advance\vm@date@height by -\dp0
    \the\year\vm@date@separator\two@digits{\month}\vm@date@separator\two@digits{\day}%
  }%
}

\newcommand{\forcewidth}{\multicolumn{3}{@{}c@{}}{\rule[-2ex]{0pt}{5ex}\rule{\insidefboxwidth}{0pt}}\\[-5ex]}

\makeatletter
\renewcommand\subparagraph{\@startsection{subparagraph}{5}{0cm}%
                                       {3.25ex \@plus1ex \@minus .2ex}%
                                       {-1em}%
                                      {\normalfont\normalsize--\hskip1ex\bfseries}}
\makeatother

\newtheorem{example}{Example}

\newlength{\insidefboxwidth}
\begin{document}

\sloppy

\maketitle

\begin{abstract}
Cypher is a query language for property graphs. It was originally designed
and implemented as part of the Neo4j graph database, and it is currently used
in a growing number of commercial systems, industrial applications and research
projects. In this work, we provide denotational semantics of the core fragment
of the read-only part of Cypher, which features in particular pattern matching,
filtering, and most relational operations on tables.
\end{abstract}

\setcounter{tocdepth}{1}
\tableofcontents
\clearpage

\section{Introduction}

In the last decade, property graph databases \cite{graphdb-intro}
such as Neo4j, JanusGraph and Sparksee have become more widespread
in industry and academia.
They have been used in multiple domains,
such as master data and knowledge management, recommendation engines, fraud detection,
IT operations and network management, authorization and access control \cite{graphdb-book},
bioinformatics~\cite{Lysenko2016}, social networks \cite{DrakopoulosKT16},
software system analysis
\cite{Hawes:2015}, and in investigative journalism~\cite{panama}.
Using graph databases to 
manage graph-structured data confers many benefits
such as explicit support for modeling graph data,
native indexing and storage for fast graph traversal operations,
built-in support for graph algorithms (e.g.,
Page Rank, subgraph matching and so on), and
the provision of graph languages, allowing users to express complex
pattern-matching operations.

This paper is about Cypher, a well-established language for
querying and updating property graph databases, which began life in
the Neo4j product, but has now been implemented commercially in other
products such as SAP HANA Graph, Redis Graph, Agens Graph (over
PostgreSQL) and Memgraph.
%
%
%
%
The data model  that is used by Cypher is that of \emph{property graphs}.
It is the most popular graph data model in
industry, and is becoming increasingly prevalent in academia \cite{LibkinMV16}.  The
model comprises \emph{nodes}, representing entities (such as people,
bank accounts, departments and so on), and \emph{relationships}
(synonymous with \emph{edges}), representing the connections or
relationships between the entities.  In the graph model, the
relationships are as important as the entities themselves.  Moreover,
any number of attributes (henceforth termed {\em properties}), in the
form of key-value pairs, may be associated with the nodes and
relationships.  This allows for the modeling and querying of complex
data.

The goal of this document is to provide denotational semantics for a core fragment of the read-only
part of Cypher, which features pattern matching, filtering, and most relational operations on tables.
Notable parts that are excluded from this work include all update (write) clauses, line-ordering
and aggregation.
Covered value types include trilean values, integers, strings, lists, maps and paths.

The need for a formal semantics stems from the fact that
Cypher, in addition to being implemented in an industrial product with
a significant customer base, has been picked up by others, and
several implementations  of it exist.
Given the lack of
a standard for the language (which can take many years to
complete, as it did for SQL), it has become pressing to agree on the
formal data model and the meaning of the main constructs.
A formal semantics has other advantages too;
for example, it allows one to reason about the equivalence of queries,
and prove correctness of existing or discover new optimizations. The
need of the formal semantics has long been accepted in the field of
programming languages \cite{mitchell-book} and for several common
languages their
semantics has been fully worked out
\cite{SMLdef,abelson-scheme,SemanticsC,Gurevich-C-semantics}.
Recently similar efforts have been made for the core SQL constructs
\cite{Veanes2010,hottsql,cosette,vldb17} with the goal of proving
correctness of SQL optimizations and understanding the expressiveness
of its features. The existence of the formal semantics of Cypher makes
it possible for different implementations to agree on its core features,
and paves a way to a reference implementation against which others
will be compared. We also note that providing semantics for an
existing real-life language like Cypher that accounts for all of its
idiosyncrasies is much harder than for theoretical calculi underlying
main features of languages, as has been witnessed by previous work on
SQL \cite{vldb17} and on many programming languages.

\medskip

%

%
The document is organized as follows.
Section~\ref{sec:principle} is an overview of the semantics.
Section~\ref{sec:model} defines the data model that
will be used throughout the document. This includes base data values that
can occur in property graphs or be returned by queries, as well as property
graphs themselves, and finally records and tables on which the semantics of queries
are based.
Section~\ref{sec:pattern} defines the core mecanism of Cypher that is, \emph{pattern matching}.
It provides the syntax of patterns, defines the notion of rigid patterns
and explicits how to compute the bag of the paths that satisfy a pattern.
Then, section~\ref{sec:syntax} provides a formal grammar that defines the syntax
of the fragment of Cypher that is considered in this work.
It is organized around the three main constructs of a Cypher statement: expressions, clauses and queries.
Finally, Section~\ref{sec:semantics} defines the semantics of Cypher over the
syntax provided in Section~\ref{sec:syntax}.
More specifically, this section defines how to evaluate an expression as a value, and
to formally specify a Cypher query as a mathematical function that returns tables of values.
It is important to note that the sole purpose of this work is to formally
specify the intended behaviour of Cypher.
It should not be considered as a user's
guide and the reader is assumed to already possess a good understanding of Cypher.
%

\section{General principles of the semantics}
\label{sec:principle}

This section provides an overview of the semantics.
Most of the object we refer to are only briefly described here.
All the proper definitions will be given later on.

\medskip

The key elements of Cypher are as
follows:
\begin{itemize}
\item data model, that includes {\em values}, {\em graphs}, and {\em tables};
\item query language, that includes {\em expressions}, {\em patterns}, {\em
    clauses}, and {\em queries}.
\end{itemize}

\emph{Values} can be simple, such as strings and integers, or composite, such as lists and maps.
Cypher is a language to query data from \emph{property graphs}.
As usual, such a graph consists of \emph{nodes} that are linked by directed
edges, called \emph{relationships} but in addition, relationships bear
\emph{types}, nodes bear \emph{labels} and both may bear properties, i.e.
key-value pairs.
%
Expressions
denote values; patterns occur in \sqlkw{MATCH} clauses; and queries
are sequences of clauses.
Tables are bags of \emph{records}, which are partial functions
from (column-)\emph{names} to values; in other words, tables
are neither line-ordered nor column-ordered.
Each clause denotes a function from tables
to tables and each query returns a table.

To provide a formal semantics of Cypher, we will define one relation and two
functions:
\begin{itemize}
\item The {\em pattern matching relation} checks if a path $p$ in a graph $G$
  satisfies a pattern $\pi$, under an assignment $u$ of values to the free
  variables of the pattern. This is written as $(p,G,u) \models \pi$.
\item The {\em semantics of expressions} associates an expression $\expr$, a
  graph $G$ and an assignment $u$ 
  with a value $\sem{\expr}_{G,u}$.
\item The {\em semantics of queries} (resp., {\em clauses}) associates a query
  $Q$ (resp., clause $C$) and a graph $G$ with a function $\sem{Q}_G$ (resp.,
  $\sem{C}_G$) that takes a table and returns a table (perhaps with more rows or
  with wider rows).
\end{itemize}

Note that the semantics of a query $Q$ is a {\em function}; thus it should not
be confused with the {\em output} of $Q$.
The evaluation of a query starts with the table containing one empty tuple,
which is then progressively changed by applying functions that provide
the semantics of  $Q$'s clauses. The composition of such functions, i.e., the
semantics of $Q$, is a
function again, which defines the output as
\begin{equation*}
\text{\sf output}(Q,G) ~=~ \sem{Q}_G(T_{\text{unit}})
\end{equation*}
where $T_{\text{unit}}$ is the table containing a single empty record.

\medskip

With this basic understanding of the data model and the semantics of the
language, we now explain it in detail. Throughout the description of the
semantics, we shall use the notational conventions in Table~\ref{tab:notation}
(they will be explained in the following sections; they are summarized here for
a convenient reference).

\begin{table}
  \centering
  \begin{tabular}{lcc}
    \toprule
    {\bf Concept} & {\bf Notation} & {\bf Set notation}\\
    \midrule
    Property keys            & $k$    & $\K$\\
    Node identifiers         & $n$    & $\N$\\
    Relationship identifiers & $r$    & $\R$\\
    Node labels              & $\ell$ & $\Lab$\\
    Relationship types       & $t$    & $\T$\\
    Names                    & $a,b$  & $\A$\\
    Base functions           & $f$    & $\F$\\
    Values                   & $v$    & $\V$\\
    Expressions              & $e$    & -- \\
    Node patterns            & $\chi$ & -- \\
    Relationship patterns    & $\rho$ & -- \\
    Path patterns            & $\pi$  & -- \\
    \bottomrule
  \end{tabular}
  \caption{Summary of notational conventions}
  \label{tab:notation}
\end{table}

\section{Data Model}
\label{sec:model}

\subsection{Values}
We consider three disjoint sets $\K$ of {\em property keys},
$\N$ of {\em node identifiers} and $\R$ of {\em relationship identifiers} (ids
for short). These sets are all assumed to be countably infinite (so we never run
out of keys and ids). For this presentation of the model, we assume two base
types: the integers $\Z$, and the type of finite strings over a finite alphabet
$\Sigma$ (this does not really affect the semantics of queries; these two types
are chosen purely for illustration purposes).

The set $\V$ of values is inductively defined as follows:
\begin{itemize}
\item Identifiers (i.e., elements of $\N$ and $\R$) are values;
\item Base types (elements of $\Z$ and $\Sigma^*$) are values;
\item $\true$, $\false$ and $\nv$ are values;
\item $\cypherlist{}$ is a value (empty list), and if $v_1,\ldots,v_m$ are
  values, for $m>0$, then $\cypherlist{v_1,\ldots,v_m}$ is a value. %
\item $\cyphermap{}$ is a value (empty map), and if $k_1,\dotsc,k_m$ are
  distinct property keys and $v_1,\dotsc,v_m$ are values, for $m > 0$, then
  $\cyphermap{(k_1,v_1),\dotsc,(k_m,v_m)}$ is a value. %
\item If $n$ is a node identifier, then $\cypherpath{n}$ is a value. If
  $n_1,\dotsc,n_m$ are node ids and $r_1,\ldots,r_{m-1}$ are relationship ids,
  for $m > 1$, then $\cypherpath{n_1,r_1,n_2,\ldots,n_{m-1},r_{m-1},n_m}$ is a
  value. We shall use shorthands $n$ and $n_1r_1n_2\ldots n_{m-1}r_{m-1}n_m$.
\end{itemize}

In the Cypher syntax, lists are $[v_1,\ldots,v_m]$ and maps
are $\{k_1:v_1,\ldots,k_m:v_m\}$; we use explicit notation for them to
make clear the distinction between the syntax and the
semantics of values.

We use the symbol ``$\cdot$'' to denote concatenation of 
paths, which is possible 
only if the first path ends in a node
where the second starts, i.e., if $p_1 = n_1r_1 \dotsb r_{j-1}n_j$ and $p_2 =
n_j r_j \dotsb r_{m-1}n_m$ then $p_1 \cdot p_2$ is $n_1r_1n_2 \dotsb
n_{m-1}r_{m-1}n_m$.

Every real-life query language will have a number of functions defined on its
values, e.g., concatenation of strings and arithmetic operations on numbers. To
model this, we assume a finite set $\F$ of predefined functions that can be
applied to values (and produce new values). The semantics is parameterized by
this set, which can be extended whenever new types and/or basic functions are
added to the language.

%
%
%
%

\subsection{Property graphs}
Let $\Lab$ and $\T$ be countable sets of node labels and relationship types,
respectively. 
A property graph is a tuple $G = \langle N,R,\src,\tgt,\iota,\lambda,\tau
\rangle$ where:
\begin{itemize}
\item $N$ is a finite subset of $\N$, whose elements are referred to as the
  \emph{nodes} of $G$.
\item $R$ is a finite subset of $\R$, whose elements are referred to as the
  \emph{relationships} of $G$.
\item $\src \colon R \rightarrow N$ is a function that maps each relationship to
  its \emph{source} node.
\item $\tgt \colon R \rightarrow N$ is a function that maps each relationship to its
  \emph{target} node.
\item $\iota \colon (N \cup R) \times \K \rightarrow \V$ is a
  function that maps a (node or relationship) identifier and a property key to
  a value.
  It is assumed that~$\iota$ is a total function but that its ``non-$\nv$ support'' is
  finite: there are only finitely many~$j\in(N \cup R)$ and~$k\in\K$ such that~$\iota(j,k)\neq\nv$.

\item $\lambda \colon N \rightarrow 2^\Lab$ is a function that maps each node id
  to a finite (possibly empty) set of labels.
\item $\tau \colon R \rightarrow \T$ is a function that maps each relationship
  identifier to a relationship type. %
\end{itemize}

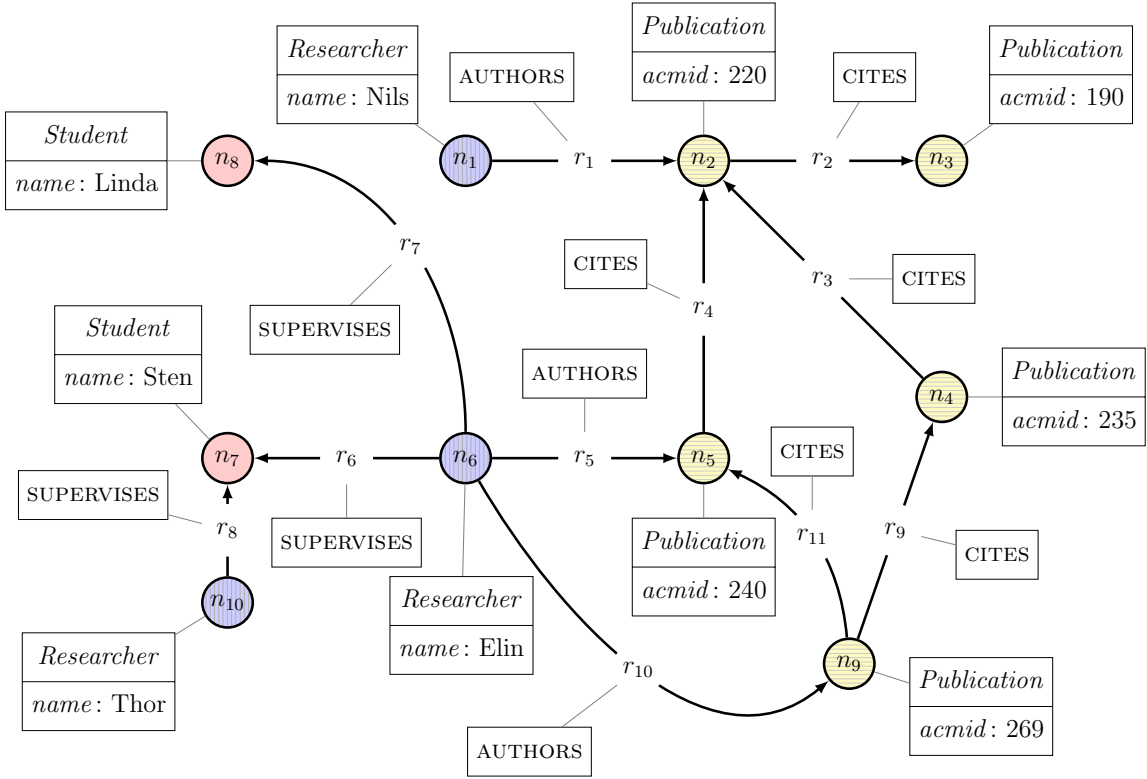
\begin{figure}
  \centering
  \resizebox{\textwidth}{!}{%
\begin{tikzpicture}[%
  node distance=4cm and 3cm,
  detail/.default={2},
  publication/.style={preaction={fill=yellow!30}, LineSep=1.6pt, pattern=hLines, pattern color=gray!40},
  researcher/.style={preaction={fill=blue!20}, LineSep=1.6pt, pattern=vLines, pattern color=gray!60},
  student/.style={fill=red!20}]
  \arraycolsep=0pt
  \node [nodeid, researcher, pin={[detail]142:%
    \strut$\mathit{Researcher}$
    \nodepart{second}
    \strut$\mathit{name}\colon\text{Nils}$
  }] (n1) {$n_1$};
  \node [nodeid, right=of n1, publication, pin={[detail]above:%
    \strut$\mathit{Publication}$
    \nodepart{second}
    \strut$\mathit{acmid}\colon\text{220}$
  }] (n2) {$n_2$};
  \node [nodeid, right=of n2, publication, pin={[detail]38:%
    \strut$\mathit{Publication}$
    \nodepart{second}
    \strut$\mathit{acmid}\colon\text{190}$
  }] (n3) {$n_3$};
  \node [nodeid, below=3cm of n3, publication, pin={[detail]right:%
    \strut$\mathit{Publication}$
    \nodepart{second}
    \strut$\mathit{acmid}\colon\text{235}$
  }] (n4) {$n_4$};
  \node [nodeid, below=of n2, publication, pin={[detail]below:%
    \strut$\mathit{Publication}$
    \nodepart{second}
    \strut$\mathit{acmid}\colon\text{240}$
  }] (n5) {$n_5$};
  \node [nodeid, below=of n1, researcher, pin={[detail, pin distance=1.5cm]-92:%
    \strut$\mathit{Researcher}$
    \nodepart{second}
    \strut$\mathit{name}\colon\text{Elin}$
  }] (n6) {$n_6$};

  \node [nodeid, left=of n6, student, pin={[detail]115:%
    \strut$\mathit{Student}$
    \nodepart{second}
    \strut$\mathit{name}\colon\text{Sten}$
  }] (n7) {$n_7$};
  \node [nodeid, above=of n7, student, pin={[detail]left:%
    \strut$\mathit{Student}$
    \nodepart{second}
    \strut$\mathit{name}\colon\text{Linda}$
  }] (n8) {$n_8$};
  \node [nodeid, below=3.5cm of n4, xshift=-1.5cm, publication, pin={[detail, yshift=-7mm]right:%
    \strut$\mathit{Publication}$
    \nodepart{second}
    \strut$\mathit{acmid}\colon\text{269}$
  }] (n9) {$n_9$};
  \node [nodeid, below=1.5cm of n7, researcher, pin={[detail]210:%
    \strut$\mathit{Researcher}$
    \nodepart{second}
    \strut$\mathit{name}\colon\text{Thor}$
  }] (n10) {$n_{10}$};

  \path [relationship] (n1) edge coordinate (@aux) (n2);
  \node [edgeid, pin={[detail=1]100:\strut\textsc{authors}}] at (@aux) {$r_1$};
  \path [relationship] (n2) edge coordinate (@aux) (n3);
  \node [edgeid, pin={[detail=1]80:\strut\textsc{cites}}] at (@aux) {$r_2$};
  \path [relationship] (n4) edge coordinate (@aux) (n2);
  \node [edgeid, pin={[detail=1, pin distance=7mm]right:\strut\textsc{cites}}] at (@aux) {$r_3$};
  \path [relationship] (n5) edge coordinate (@aux) (n2);
  \node [edgeid, pin={[detail=1]160:\strut\textsc{cites}}] at (@aux) {$r_4$};
  \path [relationship] (n6) edge coordinate (@aux) (n5);
  \node [edgeid, pin={[detail=1]90:\strut\textsc{authors}}] at (@aux) {$r_5$};

  \path [relationship] (n6) edge coordinate (@aux) (n7);
  \node [edgeid, pin={[detail=1]below:\strut\textsc{supervises}}] at (@aux) {$r_6$};
  \path [relationship] (n6) edge [out=90,in=0] coordinate (@aux) (n8);
  \node [edgeid, pin={[detail=1]-100:\strut\textsc{supervises}}] at (@aux) {$r_7$};

  \path [relationship] (n6) edge [out=-60,in=-140] coordinate (@aux) (n9);
  \node [edgeid, pin={[detail=1,xshift=-5mm]255:\strut\textsc{authors}}] at (@aux) {$r_{10}$};
  \path [relationship] (n9) edge coordinate (@aux) (n4);
  \node [edgeid, pin={[detail=1, yshift=-4mm]right:\strut\textsc{cites}}] at (@aux) {$r_9$};

  \path [relationship] (n9) edge [bend right] coordinate (@aux) (n5);
  \node [edgeid, pin={[detail=1]above:\strut\textsc{cites}}] at (@aux) {$r_{11}$};

  \path [relationship] (n10) edge coordinate (@aux) (n7);
  \node [edgeid, pin={[detail=1, yshift=0mm]170:\strut\textsc{supervises}}] at (@aux) {$r_8$};
\end{tikzpicture}%
}%
  \caption{Example data graph showing supervision and citation data for
    researchers, students and publications}
  \label{fig:data-graph}
\end{figure}

\begin{example}
  We now refer to the property graph in Figure~\ref{fig:data-graph} and show
  how, for a sample of its nodes and relationships, it is formally represented
  in this model as a graph ${G = (N,R,\src,\tgt,\iota,\lambda,\tau)}$.
  \begin{itemize}
  \item $N = \set{n_1,\ldots,n_{10}}$;
  \item $R = \set{r_1,\ldots,r_{11}}$;
  \item $\src = \left\{\; \setlength{\jot}{0pt}
      \begin{aligned}
        r_1 &\mapsto n_1 \,, & r_4 &\mapsto n_5 \,, & %
        r_7 &\mapsto n_6 \,, & r_{10} &\mapsto n_6 \\
        r_2 &\mapsto n_2 \,, & r_5 &\mapsto n_6 \,, & %
        r_8 &\mapsto n_{10} \,, & r_{11} &\mapsto n_9 \\
        r_3 &\mapsto n_4 \,, & r_6 &\mapsto n_6 \,, & %
        r_9 &\mapsto n_9 &
      \end{aligned} \;\right\}$\,;
  \item $\tgt = \left\{\; \setlength{\jot}{0pt}
      \begin{aligned}
        r_1 &\mapsto n_2 \,, & r_4 &\mapsto n_2 \,, & %
        r_7 &\mapsto n_8 \,, & r_{10} &\mapsto n_9 \\
        r_2 &\mapsto n_3 \,, & r_5 &\mapsto n_5 \,, & %
        r_8 &\mapsto n_7 \,, & r_{11} &\mapsto n_5 \\
        r_3 &\mapsto n_2 \,, & r_6 &\mapsto n_7 \,, &
        r_9 &\mapsto n_{4\phantom{9}} \phantom{\,,} &
      \end{aligned} \;\right\}$\,;
  \item $\iota(n_1,\text{name}) = \text{Nils}$, $\iota(n_2,\text{acmid}) = 220$,
    $\iota(n_3,\text{acmid}) = 190$, \ldots, $\iota(n_{10},\text{name}) =
    \text{Thor}$;
  \item $\lambda(n_1) = \lambda(n_6) = \lambda(n_{10}) = \{\text{Student}\}$,\, %
    $\lambda(n_2) = \lambda(n_3) = \lambda(n_4) = \lambda(n_5) = \lambda(n_9) =
    \{\text{Publication}\}$, %
    $\lambda(n_7) = \lambda(n_8) = \{\text{Researcher}\}$;
  \item $\tau(r) = %
    \begin{cases}
      \textsc{authors} & \text{for } r \in \{r_1,r_5,r_{10}\} \,,\\
      \textsc{supervises} & \text{for } r \in \{r_6,r_7,r_8\} \,,\\
      \textsc{cites} & \text{for } r \in \{r_2,r_3,r_4,r_9,r_{11}\} \,.
    \end{cases}$
  \end{itemize}
\end{example}


\subsection{Tables}
Let $\A$ be a countable set of names.  A $\emph{record}$ is a
partial function from names to values, conventionally denoted as a tuple with
named fields $u = (a_1 : v_1,\ldots,a_n : v_n)$ where $a_1,\ldots,a_n$ are
distinct names, and $v_1,\ldots,v_n$ are values. The order in which the fields
appear is only for notation purposes. We refer to $\dom(u)$, i.e., the domain of
$u$, as the set $\set{a_1,\ldots,a_m}$ of names used in $u$. Two records $u$ and
$u'$ are \emph{uniform} if $\dom(u)=\dom(u')$.

If $u = (a_1 : v_1,\ldots,a_n : v_n)$ and $u' = (a_1':v_1',\ldots,a_m:v_m')$ are
two records, then $(u,u')$
denotes the
record $(a_1 : v_1,\ldots,a_n : v_n,a_1':v_1',\ldots,a_m':v_m')$, assuming that
all $a_i, a_j'$ for $i\leq n, j \leq m$ are distinct.  If $A =
\set{a_1,\ldots,a_n}$ is a set of names $v$ is a value, then $(A:v)$ denotes the
record $(a_1:v, \ldots, a_n:v)$. We use $()$ to denote the empty record, i.e.,
the 
partial function from names to values whose domain is empty.

If $A$ is a set of names, then a \emph{table} with fields $A$ is a
bag, or multiset, of records $u$ such that $\dom(u)=A$. A table
with no fields
is just a bag of copies of the empty record. In most cases,
the set of fields of tables will be clear from the
context, and will not be explicitly stated. Given two tables $T$ and
$T'$, we use $T \uplus T'$ to denote their \emph{bag union}, in which
the multiplicity of each record is
the sum of their multiplicities in  $T$ and $T'$.
If $B=\{b_1,\ldots,b_n\}$ is a bag, and $T_{b_1},\ldots,T_{b_n}$ are
tables, then
$\biguplus_{b\in B} T_b$ stands for $T_{b_1}\uplus\ldots\uplus T_{b_n}$.
Finally, we use $\distinct(T)$ to denote the result of duplicate
elimination on $T$, i.e., each tuple of $T$ is present just once in
$\distinct(T)$.

\section{Pattern matching}
\label{sec:pattern}

\begin{figure}[t]%
  \renewcommand{\code}[1]{\rootcode[\color{darkred}\bfseries]{#1}}%
  \fbox{%
    \parbox{\insidefboxwidth}{%
      \abovedisplayskip=0pt\belowdisplayskip=0pt%
      \begin{align*}
        \pathpattern ::=&\phantommid \pathpattern*  \mymid a ~\code{=}~ \pathpattern*
        \\
        \pathpattern* ::=&\phantommid \nodepattern \mymid
        \nodepattern\ \relpattern\ \pathpattern*
        \\
        \nodepattern~::= &\phantommid \code{(}\, \name\qm \ \labellist\qm \ \props\qm
        \,\code{)}
        \\
        \relpattern~::=&\phantommid \code{-[}\, \name\qm \ \typelist\qm \ \iter\qm \ \props\qm \,\code{]->} \mymid \code{<-[}\, \name\qm \ \typelist\qm \ \iter\qm \ \props\qm \,\code{]-} \\[-.5mm] &
        \mymid \code{-[}\, \name\qm \ \typelist\qm \ \iter\qm \ \props\qm \,\code{]-}
        \\[2mm]
        \labellist~::= &\phantommid \code{:}\lab \mymid \code{:}\lab\ \labellist \\
        \props~::=&\phantommid \code{\{} \, \keyexprtuple \, \code{\}} \\
        \keyexprtuple ::=&\phantommid \key \code{:} \expr \mymid \key \code{:} \expr\code{,}\, \keyexprtuple \\
        \typelist~::=&\phantommid \code{:} \type  \mymid \typelist\,\code{|}\type \\
        \iter~::=&\phantommid \code{$\ast$} \mymid \code{$\ast$}d \mymid \code{$\ast$}d_1\code{..} \mymid \code{$\ast$..}d_2 \mymid \code{$\ast$}d_1\code{..}d_2 \quad\quad d,d_1,d_2 \in \Nat
      \end{align*}%
    }%
  }
  \caption{Syntax of Cypher patterns}
  \label{pattern-syntax-fig}
\end{figure}


\subsection{Syntax of patterns} It is important to remember that the Cypher grammar is defined
by mutual recursion of expressions, patterns, clauses, and queries. Here, the
description of patterns will make a reference to expressions, which we will
cover later on; all we need to know for now is that these will denote values.

The Cypher syntax of patterns is given in Figure~\ref{pattern-syntax-fig}, where
the highlighted symbols denote tokens of the language. Instead of the actual
Cypher syntax, here we use an abstract mathematical notation that lends itself
more naturally to a formal treatment.

A node pattern $\chi$ is a triple $(a,L,P)$ where:
\begin{itemize}
\item $a \in \N \cup \{\nil\}$ is an optional name;
\item $L \subset \Lab$ is a possibly empty finite set of node labels;
\item $P$ is a possibly empty finite partial map from $\K$ to expressions.
\end{itemize}
For example, the following node pattern in Cypher syntax:
\begin{center}
  \lstinline[style=cypher,mathescape]|(x:Person:Male {name: ${\normalsize \expr_1}$, age: ${\normalsize \expr_2}$})|
\end{center}
is represented as $(x,\{\text{Person},\text{Male}\},\{\text{name}\mapsto
e_1,\text{age}\mapsto e_2\})$, where $e_1$ and $e_2$ are the representations of
expressions $\expr_1$ and $\expr_2$, respectively. The simplest node
pattern \code{()} is represented by $(\nil,\emptyset,\emptyset)$.

A relationship pattern $\rho$ is a tuple $(d,a,T,P,I)$ where:
\begin{itemize}
\item $d \in \{\ltor,\rtol,\undir\}$ specifies the {\em direction} of the
  pattern: left-to-right ($\ltor$), right-to-left ($\rtol$), or undirected
  ($\undir$);
\item $a \in \N \cup \{\nil\}$ is an optional name,
\item $T \subset \T$  is a possibly empty finite set of relationship types;
\item $P$ is a possibly empty finite partial map from $\K$ to
  expressions; 
\item $I$ is either~$\nil$ or~$(m,n)$ with~$m,n\in \Nat \cup \{\nil\}$.
\end{itemize}
Table \ref{tab:example-rel-pattern} gives a few relationship patterns
and their mathematical representations.
As highlighted by these examples,~$I$ is \nil{} if and only if
the optional grammar token \iter{} does not appear in syntax of the pattern
(see Figure~\ref{pattern-syntax-fig});
otherwise,~$I$ is equal to~$(\nil,\nil)$ if~$\iter$ derives to~$\ast$ and
$I$ is equal to~$(d,d)$, $(d_1,\nil)$, $(\nil,d_2)$, $(d_1,d_2)$ if other
derivations rules are applied, respectively.
%

\begin{table}\centering
\begin{tabular}{cc}
  \toprule
  Pattern & Representation \\
  \midrule
  \lstinline[style=cypher]!-[:KNOWS {since:1985}]-!
  & $(\undir,\nil,\{\textsc{knows}\},\{\text{since}\mapsto
1985\},nil)$ \\
%
  \lstinline[style=cypher]!-[:KNOWS*1 {since:1985}]-!
  & $(\undir,\nil,\{\textsc{knows}\},\{\text{since}\mapsto
1985\},(1,1))$ \\
  \lstinline[style=cypher]!-[:KNOWS*1..1 {since:1985}]-!
  & $(\undir,\nil,\{\textsc{knows}\},\{\text{since}\mapsto
1985\},(1,1))$ \\
  \lstinline[style=cypher]!-[:KNOWS*..1 {since:1985}]-!
  & $(\undir,\nil,\{\textsc{knows}\},\{\text{since}\mapsto
1985\},(nil,1))$ \\
  \lstinline[style=cypher]!-[:KNOWS* {since:1985}]-!
  & $(\undir,\nil,\{\textsc{knows}\},\{\text{since}\mapsto
1985\},(nil,nil))$ \\
  \bottomrule
\end{tabular}
\caption{Example of relationship patterns and their representation}
\label{tab:example-rel-pattern}
\end{table}

%
In general,~$I$ defines the range of the relationship pattern.
The range is~$[m,n]$ if~$I=(m,n)$ where \nil\ is
replaced by $1$ and $\infty$ in the
place of the lower and upper bounds.
The range is~$[1,1]$ if~$I=\nil$.
A relationship pattern is said {\em rigid} if its range [m,n] satisfies:~$m=n \in \Nat$.

A path pattern is an alternating sequence of the form
\begin{equation*}
  \chi_1~~\rho_1~~\chi_2~~\dotsb~~\rho_{n-1}~~\chi_n
\end{equation*}
where each $\chi_i$ is a node pattern and each $\rho_i$ is a relationship
pattern. A path pattern $\pi$ can be optionally given a name $a$,
written as $\pi/a$; we then refer to a {\em named pattern}. A path
pattern is \emph{rigid} if all relationship patterns in it are rigid, and
\emph{variable length} otherwise.

We shall now define the satisfaction relation for path patterns w.r.t.\ a
property graph $G = (N,R,\src,\tgt,\iota,\lambda,\tau)$, a path with node ids
from $N$ and relationship ids from $R$, and an assignment $u$.

We consider rigid patterns first as a special case, because they -- unlike
variable length patterns -- uniquely define both the length and the possible
variable bindings of the paths satisfying them. The satisfaction of variable
length patterns will then be defined in terms of a set of rigid patterns.

\subsection{Satisfaction of rigid patterns}
As a precondition for a path~$p$ to satisfy any pattern (i.e.\@ for $(p,G,u) \models\pi$ to hold), it is necessary that \emph{all relationships in~$p$ are distinct}.
Then, the definition is inductive, with the
base case given by node patterns (which are trivially rigid path patterns).
Let $\chi$ be a node pattern $(a,L,P)$; then $(n,G,u) \models \chi$ if all of
the following hold:
\begin{itemize}
\item either $a$ is \nil\ or $u(a)=n$;
\item $L \subseteq \lambda(n)$;
\item $\sem{\getk{n}{k} = P(k)}_{G,u}=\true$ for each $k$ s.t.\ $P(k)$ is defined.
\end{itemize}

\begin{figure}
  \centering
  \begin{tikzpicture}[%
  node distance=2cm and 3cm,
  detail/.default={1},
  teacher/.style={preaction={fill=blue!20}, LineSep=1.6pt, pattern=vLines, pattern color=gray!60},
  student/.style={fill=red!20}]
  \arraycolsep=0pt
  \node [nodeid, teacher, pin={[detail]left:\strut$\mathit{Teacher}$}] (n1) {$n_1$};
  \node [nodeid, student, below=of n1, pin={[detail]left:\strut$\mathit{Student}$}] (n2) {$n_2$};
  \node [nodeid, teacher, right=of n2, pin={[detail]right:\strut$\mathit{Teacher}$}] (n3) {$n_3$};
  \node [nodeid, teacher, above=of n3, pin={[detail]right:\strut$\mathit{Teacher}$}] (n4) {$n_4$};

  \path [relationship] (n1) edge coordinate (@aux) (n2); 
  \node [edgeid, pin={[detail]left:\strut\textsc{knows}}] at (@aux) {$r_1$};
  \path [relationship] (n2) edge coordinate (@aux) (n3);
  \node [edgeid, pin={[detail]above:\strut\textsc{knows}}] at (@aux) {$r_2$};
  \path [relationship] (n3) edge coordinate (@aux) (n4);
  \node [edgeid, pin={[detail]right:\strut\textsc{knows}}] at (@aux) {$r_3$};
\end{tikzpicture}
  \caption{Property graph with students and teachers}
  \label{fig:teachers}
\end{figure}
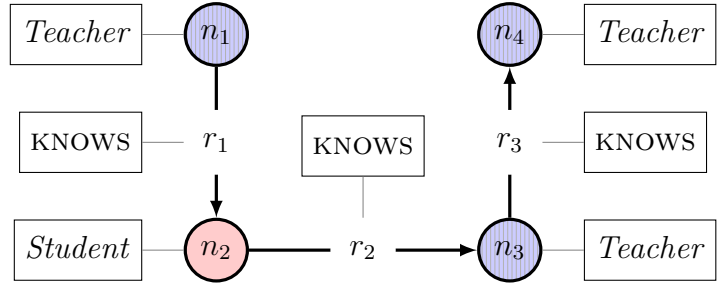

\begin{example}
  \label{exa:node-pattern}
  Consider the property graph $G$ in Figure~\ref{fig:teachers} and the node
  patterns $\chi_1 = (x,\{\text{Teacher}\},\emptyset)$ and $\chi_2 =
  (y,\emptyset,\emptyset)$. Then,
  \begin{align*}
    (n_1,G,u) &\models \chi_1 &&\text{if $u$ is an
      assignment that maps $x$ to $n_1$} \,,\\
    (n_2,G,u) &\not\models \chi_1 &&\text{for any assignment $u$} \,,\\
    (n_3,G,u) &\models \chi_1 &&\text{if $u$ is an
      assignment that maps $x$ to $n_3$} \,,\\
    (n_4,G,u) &\models \chi_1 &&\text{if $u$ is an
      assignment that maps $x$ to $n_4$} \,.
  \end{align*}
  For $i = 1,\dotsc,4$ we have that $(n_i,G,u_i) \models \chi_2$ whenever $u_i$
  is an assignment that maps $y$ to $n_i$. \hfill $\Box$
\end{example}

For the inductive case, let $\chi$ be a node pattern, let $\pi$ be a rigid
path pattern, and let $\rho$ be the relationship pattern $(d,a,T,P,I)$.
First we assume that~$ I\neq nil$, hence since~$\rho$ is rigid, the
range defined by~$I$ is~$[m,m]$ with~$m\in\N$.
For~$m = 0$, we have that $(n \cdot p,G,u) \models \chi\rho\pi$ if
\begin{enumerate}[label=(\alph{*}),topsep=1.5mm,itemsep=1mm]
\item either $a$ is \nil\ or $u(a) = \cypherlist{}$; and
\item $(n,G,u)\models \chi$ and $(p,G,u)\models\pi$.
\end{enumerate}
For $m \ge 1$, we have that $(n_1 \dotsb r_mn_{m+1} \cdot p,G,u) \models
\chi\rho\pi$ if all of the following hold:
\begin{enumerate}[label=(\alph{*}'),topsep=1.5mm,itemsep=1mm]
\item either $a$ is \nil\ or $u(a) = \cypherlist{r_1, \dotsc, r_m}$;
\item $(n_1,G,u)\models \chi$ and $(p,G,u)\models\pi$;
\end{enumerate}
and, for every $i \in \{ 1, \dotsc, m \}$, all of the following hold:
\begin{enumerate}[label=(\alph{*}'),resume,topsep=1.5mm,itemsep=1mm]
\item $\tau(r_i) \in T$;
\item $\sem{\getk{r_i}{k} = P(k)}_{G,u} = \true$ for every $k$ s.t.\ $P(k)$ is
  defined;
\item $\left(\src(r_i),\tgt(r_i)\right) \in
  \begin{cases}
    \{(n_i,n_{i+1}),(n_{i+1},n_i)\} & \text{if $d$ is } \undir \enspace ,\\
    \{(n_i,n_{i+1})\} & \text{if $d$ is } \ltor \enspace ,\\
    \{(n_{i+1},n_i)\} & \text{if $d$ is } \rtol \enspace .
  \end{cases}$
\end{enumerate}

Second, the case~$I=\nil$ is treated as if~$I=(1,1)$ with the exception that
item~(a) is replaced by:\quad\begin{enumerate*}[label=(\alph{*}'),start=1]
  \item either $a$ is \nil\ or $u(a) = r_1$
\end{enumerate*}

\begin{example}
  \label{exa:rigid-pattern}
  Consider again the property graph $G$ in Figure~\ref{fig:teachers} and the
  following rigid pattern $\pi$ in Cypher syntax:
  \begin{center}
    \lstinline[style=cypher]!(x:Teacher) -[:KNOWS*2]-> (y)!
  \end{center}
  In our mathematical representation this amounts to:
  \begin{equation*}
    \underbrace{(x,\{\text{Teacher}\},\emptyset)}_{\chi_1} \,,\, %
    \underbrace{(\rightarrow,\nil,\{\textsc{knows}\},\emptyset,(2,2))}_{\rho} \,,\, %
    \underbrace{(y,\emptyset,\emptyset)}_{\chi_2}
  \end{equation*}
  where $\chi_1$ and $\chi_2$ are the node patterns we have seen in
  Example~\ref{exa:node-pattern}. %
  Now, let $u = \{ x \mapsto n_1, y \mapsto n_3 \}$; from that example we know
  that $(n_1,G,u) \models \chi_1$ and $(n_3,G,u) \models \chi_2$. %
  Then, following the definition of satisfaction given above, one can easily see
  that $(p,G,u) \models \pi$, where $p = n_1r_1n_2r_2n_3$ and $\pi =
  \chi_1\rho\chi_2$.

  Observe that if there is another assignment $u'$ s.t.\ $(p,G,u')\models \pi$,
  then $u'$ maps $x$ to $n_1$ and $y$ to $n_3$.  This is the intuitive reason
  why rigid patterns are of interest: given a path and a rigid pattern, there
  exists at most one possible assignment of the \emph{free variables} (which we
  shall formally define shortly) of the pattern w.r.t.\ which the path
  satisfies the pattern. We will see that for variable length patterns this is
  no longer the case. \hfill $\Box$
\end{example}

For named rigid patterns, we have that $(p,G,u)\models \pi/a$ if $u(a) = p$ and
$(p,G,u)\models \pi$.

\subsection{Satisfaction of variable length patterns} %
Informally, a variable length pattern is a compact representation for a possibly
infinite set of rigid patterns; e.g., a pattern of length {\em at least} 1 will
represent patterns of length 1, patterns of length 2, and so on.

To make this idea precise, let $\rho = (d,a,T,P,(m,n))$ be a variable length
relationship pattern, and $\rho' = (d,a,T,P,(m',m'))$ be a rigid relationship
pattern.
We say that $\rho$ {\em subsumes} $\rho'$, and write $\rho \sqsupset
\rho'$, if $m'$ belongs to the range~$[m,n]$ defined by~$I$.
If~$\rho$ is rigid, then it only subsumes itself.
\OMIT{
all of the following hold:
\begin{itemize}
\item $m = 0$ or $n = 0$, whenever $m' = 0$;
\item $m \ge m'$, whenever $m \in \Nat$ and $m' \ge 1$;
\item $n \ge m'$, whenever $n \in \Nat$ and $m' \ge 1$.
\end{itemize}
} 
This subsumption relation is easily extended to path patterns. Given a variable
length pattern $\pi=\chi_1\rho_1\chi_2\dotsb\chi_{k-1}\rho_{k-1}\chi_{k}$ and a
rigid pattern $\pi'=\chi_1\rho'_1\chi_2 \dotsb \chi_{k-1}\rho'_{k-1}\chi_{k}$, we
say that $\pi$ subsumes $\pi'$ (written $\pi \sqsupset \pi'$) if $\rho_i
\sqsupset \rho'_i$ for every $i \in \{ 1, \dotsc, k-1 \}$.

Then, we define the {\em rigid extension} of $\pi$ as
\begin{equation*}
  \rigid(\pi) = \left\{\, \pi' \mid %
    \pi' \text{ is rigid and } \pi\sqsupset\pi' \,\right\} \,,
\end{equation*}
that is, the (possibly infinite) set of all rigid patterns subsumed by $\pi$.
For a named pattern, $\rigid(\pi/a) = \{\pi'/a \mid \pi' \in \rigid(\pi)\}$.
Finally, $(p,G,u) \models \pi$ if $(p,G,u) \models \pi'$ for some $\pi' \in
\rigid(\pi)$, and similarly for named patterns.

\begin{example}
  \label{exa:var-length}
  Consider the following variable length pattern $\pi$:
\begin{cypher}
(x:Teacher) -[:KNOWS*1..2]-> (z)
            -[:KNOWS*1..2]-> (y:Teacher)
\end{cypher}
  That is, $\pi$ is the pattern $\chi_1\rho\chi_2\rho\chi_3$ with
  \begin{align*}
    \chi_1 &= (x,\{\text{Teacher}\},\emptyset) \,,
    & \chi_3 &= (y,\{\text{Teacher}\},\emptyset) \,,\\
    \chi_2 &= (z,\emptyset,\emptyset) \,,
    & \rho &= (\ltor,\nil,\{\textsc{knows}\},\emptyset,(1,2)) \,.
  \end{align*}
  Then, $\rigid(\pi)$ is the set
  \begin{equation*}
    \bigl\{\, \underbrace{\chi_1\rho_1\chi_2\rho_1\chi_3}_{\pi_1}\,,\;
    \underbrace{\chi_1\rho_1\chi_2\rho_2\chi_3}_{\pi_2}\,,\;
    \underbrace{\chi_1\rho_2\chi_1\rho_1\chi_3}_{\pi_3}\,,\;
    \underbrace{\chi_1\rho_2\chi_2\rho_2\chi_3}_{\pi_4} \,\bigr\}
  \end{equation*}
  where
  \begin{equation*}
    \rho_1 = (\ltor,\nil,\{\textsc{knows}\},\emptyset,(1,1))\,, \;
    \rho_2 = (\ltor,\nil,\{\textsc{knows}\},\emptyset,(2,2))\,.
  \end{equation*}

  Consider again the property graph $G$ in Figure~\ref{fig:teachers}. Let
  \begin{align*}
    p_1 &= n_1r_1n_2r_2n_3 %
    & u_1 &= \{\,{x \mapsto n_1}, {y \mapsto n_3}, {z \mapsto n_2}\,\}\\
    p_2 &= n_1r_1n_2r_2n_3r_3n_4 %
    & u_2 &= \{\,{x \mapsto n_1}, {y \mapsto n_4}, {z \mapsto n_2}\,\}
  \end{align*}
  Then, $(p_1,G,u_1) \models \pi_1$ and $(p_2,G,u_2) \models \pi_2$; therefore,
  $\pi$ is satisfied in $G$ by $p_1$ under $u_1$ and by $p_2$ under $u_2$.  This
  shows the ability of a variable length pattern to match paths of varying
  length.

  In addition, variable length patterns may admit several assignments even for a
  single given path. To see this, note that $p_2$ satisfies $\pi$ in $G$ {\em
    also} under the assignment $u'_2$ that agrees with $u_2$ on $x$ and $y$ but
  maps $z$ to $n_3$, because $(p_2,G,u'_2) \models \pi_3$.
  \hfill $\Box$
\end{example}

In Cypher, we want to return the ``matches'' for a pattern in a graph, not
simply check whether the pattern is satisfied (i.e., there exists a match). This
is captured formally next.

\subsection{Pattern matching} %
The set of {\em free variables} of a node pattern $\chi=(a,L,P)$, denoted by
$\freevars(\chi)$, is $\{a\}$ whenever $a$ is not $\nil$, and empty
otherwise. For a relationship pattern $\rho$, the set $\freevars(\rho)$ is
defined analogously.  Then, for a path pattern $\pi$ we define $\freevars(\pi)$
to be union of all free variables of each node and relationship pattern
occurring in it.  For example, for the pattern $\pi$ of
Example~\ref{exa:var-length} we have $\freevars(\pi) = \{x,y,z\}$. %
For named patterns, $\freevars(\pi/a)= \freevars(\pi) \cup \{a\}$.
Then, for a path pattern $\pi$ (optionally named),
a graph $G$ and
an assignment $u$, we define
\begin{equation}
  \label{match-eq}
  \match(\pi,G,u) ~=~
  \biguplus_{%
    \mathclap{\substack{%
        p \text{ in } G\\
        \pi' \in \rigid(\pi)}}}
  \;\,\left\{\, u' \,\left|\!
      \begin{array}{l}
        \dom(u')=\freevars(\pi)-\dom(u)\\
        \text{and } (p, G, u \cdot u') \models \pi'
      \end{array}
    \right.\!\!\!\right\}
\end{equation}
Note that, even though both $u'$ and $\pi'$ range over infinite sets, only a
finite number of values contribute to a non-empty set in the final union. Thus
$\match(\pi,G,u)$ is finite.

In \eqref{match-eq}, $\biguplus$ stands for bag union: whenever a new
combination of $\pi'$ and $p$ is found such that $(p,G,{u \cdot u'}) \models
\pi'$, a new occurrence of $u'$ is added to $\match(\pi,G,u)$. This is in line
with the way Cypher combines the $\kwcode{MATCH}$ clause and bag semantics,
which is not captured by the satisfaction relation alone.

\begin{example}
  \label{exa:mult-match}
  Consider once again the graph $G$ in Figure~\ref{fig:teachers}, and let $\pi$
  be the following variable length pattern:
\begin{cypher}
(x:Teacher) -[:KNOWS*1..2]-> ()
            -[:KNOWS*1..2]-> (y:Teacher)
\end{cypher}
  This is similar to the pattern in Example~\ref{exa:var-length}, but
 the middle node pattern is not given any name here:
  $\freevars(\pi) = \{x,y\}$. Indeed, $\rigid(\pi)$ is the same as in the
  previous example, with $\chi_2 = (\nil,\emptyset,\emptyset)$.

  Let $p = n_1r_1n_2r_2n_3r_3n_4$ and $u = \{{x \mapsto n_1},{y \mapsto n_4}\}$;
  it is easy to see that $(p,G,u) \models \pi_3 \in \rigid(\pi)$. However,
  observe that $(p,G,u)\models \pi_2$ as well (whereas $\pi_1$ and $\pi_4$ are
  not satisfied by any path of $G$). This shows that there may be multiple ways
  for a single path to satisfy a variable length pattern even under the same
  assignment. 
In our example, two
  copies of $u$ will be added to $\match(\pi,G,\emptyset)$.  \hfill $\Box$
\end{example}


\subsection{Matching tuples of path patterns} %
Cypher allows one to match a tuple $\bar \pi = (\pi_1,\ldots,\pi_n)$ of path
patterns, each optionally named. We say that $\bar{\pi}$ is rigid if all its
components are rigid, and $\rigid(\bar{\pi})$ is defined as $\rigid(\pi_1)
\times \dotsb \times\rigid(\pi_n)$. The set of free variables of $\bar\pi$ is
defined as $\freevars(\bar{\pi}) = \bigcup_{\pi_i}{\freevars(\pi_i)}$.
Let $\bar{p} = (p_1,\dotsc,p_n)$ be a tuple of paths;
we write $(\bar p, G, u) \models \bar \pi$ if
{\em no relationship id
occurs in more than one path in} $\bar p$ and $(p_i,G,u)\models \pi_i$
for each $i \in \{1,\dotsc,n\}$.
Then, for a tuple of patterns $\bar{\pi}$, a graph $G$ and an assignment $u$,
$\match(\bar\pi,G,u)$ is defined as in \eqref{match-eq}, with the difference
that the bag union is now over {tuples} $\bar\pi'\in\rigid(\pi)$
and $\bar p$ of paths.

\section{Complete Syntax}
\label{sec:syntax}

We now present the key components of Cypher, namely {\em expressions}, {\em
  clauses}, and {\em queries}, and define their formal semantics. Together with
pattern matching defined in the previous section, they will constitute the
formalization of the core of Cypher.


%
The syntax of Cypher patterns was given in Figure~\ref{pattern-syntax-fig}.
\emph{Expressions} derives from the token \expr, whose derivation rules are shown
in Figure~\ref{fig:cypher-syntax-expr}.
Similarly, \emph{queries} derive from the token \query{}
(Figure~\ref{fig:cypher-syntax-query}) and \emph{clauses} from the token \clause{}
(Figure~\ref{fig:cypher-syntax-clause}).
%

%
%

\begin{figure}[h]%
\abovedisplayskip=0pt\belowdisplayskip=0pt%
\renewcommand{\arraystretch}{1.3}
\fbox{%
  \begin{minipage}{\insidefboxwidth}%
    \renewcommand{\code}[1]{\rootcode[\color{red}\bfseries]{#1}}%
  \begin{gather*}%
\begin{array}{r@{}l@{}r}%
  \forcewidth
  \expr~::=& \phantommid v \mymid a \mymid f\code{(} \exprlist?
  \code{)} \ \ \ \ \ \ \ \ v\in\V, \ a\in\A, \ f\in\F & \hspace*{-3cm}values/variables \\
          &\mymid \expr\code{.}k \mymid \code{\{\}} \mymid \code{\{}\,
  \keyexprtuple \,\code{\}} & maps \\
          &\mymid \code{[}\,\code{]} \mymid \code{[}\, \exprlist \,\code{]}
          \mymid \expr \ \kwcode{IN} \ \expr \\
          & \mymid
          \expr\code{[}\expr\code{]} \mymid
          \expr\code{[}\expr\code{..]} \mymid
          \expr\code{[..}\expr\code{]} \mymid
          \expr\code{[}\expr\code{..}\expr\code{]}   & \multirow{-2}{*}{$lists$}\\
          &\mymid \expr \ \kwcode{STARTS\code{\char32}WITH} \ \expr
           \mymid \expr \ \kwcode{ENDS\code{\char32}WITH} \ \expr
          & \multirow{2}{*}{$strings$}\\
          &\mymid \expr \ \kwcode{CONTAINS} \ \expr \\
          &\mymid \expr \ \kwcode{OR} \ \expr \mymid \expr \ \kwcode{AND} \ \expr \mymid \expr \ \kwcode{XOR} \ \expr \mymid \kwcode{NOT} \ \expr
          \\&\mymid \expr \ \kwcode{IS NULL} \mymid \expr \ \kwcode{IS NOT
            NULL} & \multirow{-2}{*}{$logic$}\\
          &\mymid \expr \ \code{<}\  \expr \mymid \expr \ \code{<=}\  \expr \mymid \expr \ \code{>=}\  \expr \mymid \expr \ \code{>}\  \expr
          \\& \mymid \expr \ \code{=}\  \expr \mymid \expr
          \ \code{<>}\  \expr & \hspace*{-5cm}\multirow{-2}{*}{$comparison$}
\end{array}\\%
\begin{array}{r@{}l@{}r}%
\forcewidth
\exprlist~::=&  \phantommid \expr \mymid \expr\code{,}\,\exprlist &
  expression\ lists
\end{array}%
\end{gather*}%
\end{minipage}%
}
\caption{Syntax of expressions}
\label{fig:cypher-syntax-expr}
\end{figure}

\begin{figure}[h]%
\abovedisplayskip=0pt\belowdisplayskip=0pt%
\renewcommand{\arraystretch}{1.3}
\fbox{%
  \begin{minipage}{\insidefboxwidth}%
    \renewcommand{\code}[1]{\rootcode[\color{red}\bfseries]{#1}}
  \begin{gather*}%
  \begin{array}{r@{}l@{}r}
  \forcewidth
          \query ::=& \phantommid \query* \mymid \query \ \kwcode{UNION}\ [\kwcode{ALL}]
          &
          unions \\
          \query* ::=& \phantommid \kwcode{RETURN}\ \retarg \mymid
          \clause\ \query* & \hspace*{-5mm}clause~sequences \\
  \end{array}\\%
  \begin{array}{r@{}l@{}rrrrr}
  \forcewidth
  \retarg ::=&  \phantommid \code{$\ast$} \mymid \aggexpr \ [\kwcode{AS} \ a]  \mymid
  \mymid \retarg\,\code{,}\ \aggexpr \ [\kwcode{AS} \ a]  & return\ lists \\
  \end{array}
\end{gather*}%
\end{minipage}%
}
\caption{Syntax of queries}
\label{fig:cypher-syntax-query}
\end{figure}

\begin{figure}[h]%
\abovedisplayskip=0pt\belowdisplayskip=0pt%
\renewcommand{\arraystretch}{1.3}
\fbox{%
  \begin{minipage}{\insidefboxwidth}%
    \renewcommand{\code}[1]{\rootcode[\color{red}\bfseries]{#1}}
  \begin{gather*}%
\begin{array}{r@{}l@{}r}
\forcewidth
\clause~::=& \phantommid
[\kwcode{OPTIONAL}]\ \kwcode{MATCH}\ \pathpatternlist\ [\kwcode{WHERE}\ \expr]
& matching\ clauses \\
        &\mymid \kwcode{WITH}\ \retarg\ [\kwcode{WHERE}\ \expr] \\
        & \mymid \kwcode{UNWIND} \ \expr \ \kwcode{AS} \ a \quad
        \ \ \ \  a \in  \A & relational\ clauses\\
\end{array}\\
\begin{array}{r@{}l@{}r}
\forcewidth
\pathpatternlist~ ::=& \phantommid \pathpattern \mymid \pathpattern
\code{,}\,\pathpatternlist & tuples\ of\ patterns
\end{array}
\end{gather*}%
\end{minipage}%
}
\caption{Syntax of clauses}
\label{fig:cypher-syntax-clause}
\end{figure}

\section{Complete Semantics}
\label{sec:semantics}

\subsection{Semantics of expressions}
The semantics of an expression e{} is a value $\sem{e}_{G,u}$ in $\V$
determined by a property graph $G$ and an assignment $u$ that provides bindings
for the names used in e.
The rules here are fairly straightforward and given in details below.

\medskip

Assume that we are given a fixed property graph
${G = (N,R,s,t,\iota,\lambda,\tau)}$ and a fixed record $u = (a_1 : v_1,\ldots,a_n : v_n)$ that
associates values $v_1,\ldots,v_n$ with names $a_1,\ldots,a_n$.

\paragraph{Values and variables}

\begin{itemize}[itemsep=2mm]
	\item $\sem{v}_{G,u} = v$ \\
	where $v$ is a value.
	\item $\sem{a}_{G,u} = u(a)$ \\
	where $a$ is a name that belongs to the domain of $u$.
	\item $\sem{f(e_1,\ldots,e_m)}_{G,u} = f(\sem{e_1}_{G,u},\ldots,\sem{e_m}_{G,u})$\\
	where
		$e_1,\ldots,e_m$ are expressions, and $f$ is any
          $m$-ary function in $\F$ from values to
		values.
\end{itemize}

\paragraph{Maps}
\begin{itemize}[itemsep=2mm]
	\item $\sem{e.k}_{G,u} = \left \{ \begin{array}{lr@{}l}
													\iota(\sem{e}_{G,u},k) & \text{if }& \sem{e}_{G,u} \in \N \cup \R \\
													w_i & \text{if }& \sem{e}_{G,u} = \cyphermap{(k_1,w_1),(k_2,w_2),\ldots, (k_m,w_m)} \\ && \text{ and } k=k_i \\
													\nv & \text{if }& \sem{e}_{G,u} = \cyphermap{(k_1,w_1),(k_2,w_2),\ldots, (k_m,w_m)} \\ && \text{and } k\notin\set{k_1,\ldots,k_m} \\
														& \text{or }& \sem{e}_{G,u} = \set{} \\
														& \text{or }& \sem{e}_{G,u} = \nv
												\end{array} \right.$

		where $k$ and the $k_i$'s are property keys, and $w$ and the $w_i$'s are values.

	\item $\sem{\set{k_1:e_1,\ldots,k_m:e_m}}_{G,u} = \text{map}((k_1,\sem{e_1}_{G,u}),\ldots,(k_m,\sem{e_m}_{G,u}))$

		where $k_1,\ldots,k_m$ are \textbf{distinct} property keys and $e_1,\ldots,e_m$ are expressions.

	\item $\sem{\set{k_1:e_1,\ldots,k_m:e_m}}_{G,u} = \sem{\set{k_{i_1}:e_{i_1},\ldots,k_{i_\ell}:e_{i_\ell}}}_{G,u}$

		where $k_1,\ldots,k_m$ are property keys, $e_1,\ldots,e_m$ are expressions, and
		$i_1,\ldots,i_\ell$ are distinct indices such that $\set{k_{i_1},\ldots,k_{i_\ell}} = \set{k_1,\ldots,k_m}$
		and for each $p$ such that $i_p < m$, $k_{i_p} \notin \set{k_{i_p+1},\ldots,k_m}$. In other
		words, if there are repeated keys among $k_1,\ldots,k_m$, only the last occurrence of each key
		is kept.

	\item $\sem{\set{}}_{G,u} = \text{map}()$
\end{itemize}

\paragraph{Explicit Lists}

\begin{itemize}[itemsep=2mm]
	\item $\sem{\symbollist{e_1,\ldots,e_m}}_{G,u} = \cypherlist{\sem{e_1}_{G,u},\ldots,\sem{e_m}_{G,u}}$

		where $e_1,\ldots,e_m$ are expressions.
	\item $\sem{\symbollist{~}}_{G,u} = \cypherlist{}$
\end{itemize}

\paragraph{Operations on non-empty lists} Assume that $e$ is an expression such that
$\sem{e}_{G,u} = \cypherlist{w_0,\ldots,w_{m-1}}$ for some values
$w_0,\ldots,w_{m-1}$. Then the semantics of list expressions is as
follows.

\begin{itemize}[itemsep=2mm]
	\item $\sem{e[e']}_{G,u} = \left \{ \begin{array}{lll}
														w_{i} & \text{ if } 0\leq i < m \\
														w_{m+i} & \text{ if } -m \leq i < 0 \\
														\nv & \text{ if } i < -m \text{ or } i \geq m
													\end{array} \right.$

		where $\sem{e'}_{G,u} = i$, for some integer $i$.

	\item $\sem{e[e_1..e_2]}_{G,u} = \left \{ \begin{array}{lll}
														\cypherlist{w_{\text{max}(0,i')},\ldots,w_{\text{min}(m-1,j'-1)}} & \text{ if } i' \leq j', i' < m, j' > 0 \\
														{\cypherlist{}} & \text{ otherwise}
													\end{array} \right.$

		where $\sem{e_1}_{G,u} = i$ for some integer $i$, $\sem{e_2}_{G,u} = j$ for some integer $j$,
		$i' = i$ if $i \geq 0$ and $i' = m+i$ otherwise, $j' = j$ if $j \geq 0$ and $j' = m+j$
		otherwise.

	\item $\sem{e[e_1..]}_{G,u} = \sem{e[e_1..m]}_{G,u}$

	\item $\sem{e[..e_2]}_{G,u} = \sem{e[0..e_2]}_{G,u}$

	\item $\sem{e'\ \kwcode{IN}\ e}_{G,u} = \left \{ \begin{array}{lll}
														\true & \text{if } ~\sem{e' = w_i }_{G,u} = \true~, \\ & \text{for some integer }i,~0\leq i < n \\
														\nv & \text{if the previous case does not hold } \\
														    & \text{and }  ~\sem{e' = w_i }_{G,u} = \nv~,\\& \text{for some integer }i,~0\leq i < n  \\
														\false & \text{otherwise}
													\end{array} \right.$
\end{itemize}

\paragraph{Operations on empty lists} Assume that $e$ is an expression such that
$\sem{e}_{G,u} = \cypherlist{}$.  Then the semantics of list expressions is as
follows.

\begin{itemize}[itemsep=2mm]

	\item $\sem{e\code{[}e'\code{]}}_{G,u} = \nv$\\
          where $\sem{e'}_{G,u} = i$ for some integer $i$.

	\item $\sem{e\code{[}e_1\code{..}e_2\code{]}}_{G,u} = {\cypherlist{}}$

		where $\sem{e_1}_{G,u} = i$ for some integer $i$ and $\sem{e_2}_{G,u} = j$ for some integer $j$.

	\item $\sem{e\code{[}e_1\code{..]}}_{G,u} = {\cypherlist{}}$

		where $\sem{e_1}_{G,u} = i$ for some integer $i$.

	\item $\sem{e\code{[..}e_2\code{]}}_{G,u} = {\cypherlist{}}$

		where $\sem{e_2}_{G,u} = j$ for some integer $j$.

	\item $\sem{e'\ \code{IN}\ e}_{G,u} = \false$ \\
	      where~$\sem{e'}_{G,u}$ is defined.

\end{itemize}

\paragraph{Strings} Assume that $e$ and~$e'$ are expressions such that~$\sem{e}_{G,u}$
and~$\sem{e'}_{G,u}$ belong to~$\Sigma^* \cup\{\nv\}$.

\begin{itemize}[itemsep=2mm]
	\item $\sem{e \ \kwcode{STARTS WITH}\ e'}_{G,u} =
		\left \{ \begin{array}{lll}
			\true & \text{if }\exists s, \sem{e}_{G,u} = \sem{e'}_{G,u}\cdot s \\[1.5mm]
			\nv & \text{if } \sem{e}_{G,u} = \nv \\&\text{or } \sem{e'}_{G,u} = \nv \\[1.5mm]
			\false & \text{otherwise}
		\end{array} \right.$
%

	\item $\sem{e \ \kwcode{ENDS WITH}\ e'}_{G,u} =
		\left \{ \begin{array}{lll}
			\true & \text{if }\exists s, \sem{e}_{G,u} = s \cdot \sem{e'}_{G,u} \\[1.5mm]
			\nv & \text{if } \sem{e}_{G,u} = \nv \\&\text{or } \sem{e'}_{G,u} = \nv \\[1.5mm]
			\false & \text{otherwise}
		\end{array} \right.$
%

	\item $\sem{e \ \kwcode{CONTAINS}\ e'}_{G,u} =
		\left \{ \begin{array}{lll}
			\true & \text{if }\sem{e}_{G,u} = s_1 \cdot \sem{e'}_{G,u} \cdot s_2
			\\& \text{for some strings }s_1,s_2
			\\[1.5mm]
			\nv & \text{if } \sem{e}_{G,u} = \nv\\& \text{or } \sem{e'}_{G,u} = \nv \\[1.5mm]
			\false & \text{otherwise}
		\end{array} \right.$
%
\end{itemize}

\paragraph{Logic}

Assume that $e$ and $e'$ are expressions such that~$\sem{e}_{G,u}$
and~$\sem{e'}_{G,u}$ both belong to~$\set{\true,\false,\nv}$.

\begin{itemize}[itemsep=2mm]
	\item $\sem{e\ \kwcode{OR}\ e'}_{G,u} = \left \{ \begin{array}{lll}
														\true & \text{ if } \sem{e}_{G,u} = \true \text{ or }  \sem{e'}_{G,u} = \true\\
														\false & \text{ if } \sem{e}_{G,u} = \sem{e'}_{G,u} = \false\\
														\nv & \text{ otherwise}
													\end{array} \right.$
	\item $\sem{e\ \kwcode{AND}\ e'}_{G,u} = \left \{ \begin{array}{lll}
														\true & \text{ if } \sem{e}_{G,u} = \sem{e'}_{G,u} = \true \\
														\false & \text{ if } \sem{e}_{G,u} = \false \text{ or }  \sem{e'}_{G,u} = \false\\
														\nv & \text{ otherwise}
													\end{array} \right.$
	\item $\sem{e\ \kwcode{XOR}\ e'}_{G,u} = \left \{ \begin{array}{lll}
														\nv & \text{ if } \sem{e}_{G,u} = \nv \text{ or }  \sem{e'}_{G,u} = \nv\\
														\false & \text{ if } \sem{e}_{G,u} = \sem{e'}_{G,u} \text{ and } \sem{e}_{G,u} \neq \nv\\
														\true & \text{ otherwise} \\
													\end{array} \right.$
	\item $\sem{\kwcode{NOT}\ e}_{G,u} = \left \{ \begin{array}{lll}
														\true & \text{ if } \sem{e}_{G,u} = \false \\
														\false & \text{ if } \sem{e}_{G,u} = \true \\
														\nv & \text{ if } \sem{e}_{G,u} = \nv \\
													\end{array} \right.$
\end{itemize}

\paragraph{Value Comparisons}

\subparagraph{Nulls} The rules follow SQL: in an expression,
if an argument is \nv, then the value of the expression is \nv. The
semantics of \kwcode{IS NULL} is also the same as for SQL.

\begin{itemize}[itemsep=2mm]
	\item $\sem{e \star e'}_{G,u} = \nv$ if either $\sem{e}_{G,u} = \nv$ or $\sem{e'}_{G,u} = \nv$,
		for ${\star \in \set{<,<=,>=,>,=,<>}}$.

	\item $\sem{e \ \kwcode{IS NULL}}_{G,u} = \left \{ \begin{array}{lll}
														\true & \text{ if } \sem{e}_{G,u} = \nv \\
														\false & \text{ if } \sem{e}_{G,u} \neq \nv \\
													\end{array} \right.$

	\item $\sem{e \ \kwcode{IS NOT NULL}}_{G,u} = \left \{ \begin{array}{lll}
														\true & \text{ if } \sem{e}_{G,u} \neq \nv \\
														\false & \text{ if } \sem{e}_{G,u} = \nv \\
													\end{array} \right.$
\end{itemize}

\subparagraph{Base Types} Assume that both $e$ and $e'$ are expressions
such that $\sem{e}_{G,u}$  and $\sem{e'}_{G,u}$ are of the same base type.

\begin{itemize}[itemsep=2mm]
	\item $\sem{e = e'}_{G,u} =
		\left \{ \begin{array}{ll}
			\true & \text{if } \sem{e}_{G,u} = \sem{e'}_{G,u}\\
			\false & {\text{otherwise}}
		\end{array} \right.$
\end{itemize}

\subparagraph{Identifiers} Assume that both $e$ and $e'$ are expressions
such that $\sem{e}_{G,u}$  and $\sem{e'}_{G,u}$ are both node identifiers
or both relationship identifiers.

\begin{itemize}[itemsep=2mm]
	\item $\sem{e = e'}_{G,u} =
		\left \{ \begin{array}{ll}
			\true & \text{if } \sem{e}_{G,u} = \sem{e'}_{G,u}\\
			\false & {\text{otherwise}}
		\end{array} \right.$
\end{itemize}

\subparagraph{Empty maps}

Assume that both $e$ and $e'$ are expressions such that both
$\sem{e}_{G,u}$  and $\sem{e'}_{G,u}$ are maps, and one of
them is $\cyphermap{}$.

\begin{itemize}[itemsep=2mm]
	\item $\sem{e = e'}_{G,u} =
		\left \{ \begin{array}{ll}
			\true & \text{if } \sem{e}_{G,u} = \sem{e'}_{G,u} = \cyphermap{} \\
			\false & {\text{otherwise}}
		\end{array} \right.$
\end{itemize}

\subparagraph{Non-empty maps, same number of keys}

Assume that  $\sem{e}_{G,u} = \symbolmap{k_1:w_1,\ldots,k_m:w_m}$ and
${\sem{e'}_{G,u} = \symbolmap{k'_1:w'_1,\ldots,k'_m:w'_m}}$, where $k_1,\ldots,k_m,k'_1,\ldots,k'_m$ are
keys, and $w_1,\ldots,w_m,w'_1,\ldots,w'_m$ are values, and $m \geq 1$.

\begin{itemize}[itemsep=2mm]
	\item $\sem{e = e'}_{G,u} =
		\left \{ \begin{array}{lll}
			\true & \text{if} & \set{k_1,\ldots,k_m} = \set{k'_1,\ldots,k'_m} \\
					& & \text{ and }
                        \sem{e.k_i = e'.k_i}_{G,u} = \true
                        \text{ for all }i\leq m\\
			\nv & \text{if} & \set{k_1,\ldots,k_m} = \set{k'_1,\ldots,k'_m} \\
				 & & \text{and } \sem{e.k_i =
                          e'.k_i}_{G,u} = \nv \text{ for some
                        }i\leq m\\
				 & & \text{and } \sem{e.k_i =
                          e'.k_i}_{G,u} \neq\ \false \text{ for
                          all }i \leq m \\
			\false & \multicolumn{2}{l}{\text{otherwise}}
		\end{array} \right.$
\end{itemize}

\subparagraph{Non-empty maps, different number of keys}

Assume that  $\sem{e}_{G,u} = \symbolmap{k_1:w_1,\ldots,k_m:w_m}$ and
${\sem{e'}_{G,u} = \symbolmap{k'_1:w'_1,\ldots,k'_l:w'_l}}$, where $k_1,\ldots,k_m,k'_1,\ldots,k'_l$ are
keys, and $w_1,\ldots,w_m,w'_1,\ldots,w'_l$ are values, $m,l \geq 1$,
and $m \neq l$. In this case,  $\sem{e = e'}_{G,u} = \false$.

\subparagraph{Lists}

Assume that both $e$ and $e'$ are expressions such that both $\sem{e}_{G,u}$  and $\sem{e'}_{G,u}$
are list values.

\begin{itemize}[itemsep=2mm]
	\item $\sem{e = e'}_{G,u} = \true$ \\$\text{where } \sem{e}_{G,u} = \sem{e'}_{G,u} = \cypherlist{}$
	\item $\sem{e = e'}_{G,u} = \false$ \\ $\text{where } \sem{e}_{G,u} = \cypherlist{w_1,\ldots,w_n} \text{ and }\sem{e'}_{G,u} = \cypherlist{w'_1,\ldots,w'_m}$ and~$n\neq m$.
        \item $ \sem{e = e'}_{G,u} =
		\left \{ \begin{array}{lll}
			\true & \text{if} & \forall i, \sem{w_i = w'_i}_{G,u} = \true \\
			\nv & \text{if} &\text{the previous case does not hold } \\
				&  & \text{and } \forall i, \sem{w_i = w'_i}_{G,u} \in\set{\nv,\true}\\
			\false & \multicolumn{2}{l}{\text{otherwise}}
		\end{array} \right.$

	where $\sem{e}_{G,u} = \cypherlist{w_1,\ldots,w_m} \text{ , }\sem{e'}_{G,u} = \cypherlist{w'_1,\ldots,w'_m}$ and $w_1,\ldots,w_m,w'_1,\ldots,w'_m$ are values.
\end{itemize}

\subparagraph{Paths}

Assume that both $e$ and $e'$ are expressions such that  both $\sem{e}_{G,u}$  and $\sem{e'}_{G,u}$
are path values.

\begin{itemize}[itemsep=2mm]
	\item $\sem{e = e'}_{G,u} =
		\left \{ \begin{array}{lll}
			\true & \text{if } \sem{e}_{G,u} = \sem{e'}_{G,u} \\
			\false & \text{otherwise}
		\end{array} \right.$
\end{itemize}

\subparagraph{Mismatched composite types}

If $e$ and $e'$ are expressions such that $\sem{e}_{G,u}$ is a value of a composite type (map, list, path) and $\sem{e'}_{G,u}$ is a non-null value of a different type, then $\sem{e = e'}_{G,u} = \false$. Conversely, if $\sem{e'}_{G,u}$ is of a composite type and $\sem{e}_{G,u}$ is a non-null value of a different type, then $\sem{e = e'}_{G,u} = \false$.

\subparagraph{Base types}

If $e$ and $e'$ are expressions such that $\sem{e}_{G,u}$ and $\sem{e'}_{G,u}$ are non-null values of a non-composite type,
then $\sem{e \star e'}_{G,u}$ is allowed to be
implementation-dependent, for ${\star \in \set{<,<=,>=,>}}$. That
is, for base types implementations have freedom when it comes to
defining ordering. It is assumed however that for types considered
here (numerical and strings), these are fixed and have their standard
interpretation as ordering on numbers, and lexicographic ordering for
strings.

\subsection{Semantics of queries}
A query is either a sequence of clauses ending with the \kwcode{RETURN}
statement, or a union (set of bag) of two queries. The \kwcode{RETURN} statement
contains the return list, which is either \code{$\ast$}, or a sequence of
expressions, optionally followed by \kwcode{AS} $a$, to provide their names.

To provide the semantics of queries, we assume that there exists an
(implementation-dependent) injective function $\alpha$ that maps expressions to
names. Recall that the semantics of both queries and clauses, relative to a
property graph $G$, is a function from tables to tables, so we shall describe
its value on a table $T$, i.e., $\sem{\query}_G(T)$.


\paragraph{Return} We make the following assumptions. First, the fields of $T$ are
$b_1,\ldots,b_q$. Second, if we have a return list $e_1 \ [\kwcode{AS}\ a_1],
\ldots, e_m \ [\kwcode{AS}\ a_m]$ with optional \kwcode{AS} for some of the
expressions, then $a_i'=a_i$ if $\kwcode{AS}\ a_i$ is present in the list, and
$a_i'=\alpha(e_i)$ otherwise, with the added requirement that all the $a'_i$s
are distinct. In some rules for the semantics, some \kwcode{AS} could be
optional. It is assumed that when such optional \kwcode{AS} is present on the left
side, then it is also present on the right hand side.

\begin{itemize}[itemsep=2mm]
  \item $\sem{\kwcode{RETURN}\ *}_G(T)  = T \text{ if }T\text{ has at least one
  field}$
  \item $\sem{\kwcode{RETURN}\ *, e_1 \ [\kwcode{AS}\ a_1], \ldots, e_m
  \ [\kwcode{AS}\ a_m]}_G(T) =$\\\hspace*{0cm plus 1fill}$ \sem{\kwcode{RETURN}\ b_1 \ \kwcode{AS}\ b_1,\ldots,b_q \ \kwcode{AS}\ b_q, e_1 \ [\kwcode{AS}\ a_1], \ldots, e_m \ [\kwcode{AS}\ a_m]}_G(T)$
  \item $\sem{\kwcode{RETURN}\ e_1 \ [\kwcode{AS}\ a_1], \ldots, e_m
  \ [\kwcode{AS}\ a_m]}_G(T) =$\\\hspace*{0cm plus 1fill}$\displaystyle
		 \biguplus_{u\in T}  \left\{(a'_1 : \sem{e_1}_{G,u},
                   \ldots, a'_m : \sem{e_m}_{G,u})\right\} $
\end{itemize}

\paragraph{Union } Let~$Q_1, Q_2$ be queries.

\begin{itemize}[itemsep=2mm]
  \item $\sem{Q_1 \ \kwcode{UNION ALL} \ Q_2}_G(T) = \sem{Q_1}_G(T) \cup \sem{Q_2}_G(T)$

  \item $\sem{Q_1 \ \kwcode{UNION} \ Q_2}_G(T) = \distinct\big(\sem{Q_1}_G(T) \cup \sem{Q_2}_G(T)\big)$ \\
  (Recall that $\distinct$ is the function computing duplicate elimination.)
\end{itemize}

\paragraph{Clause list}

\begin{itemize}
  \item $\sem{C\ Q}_G(T) =  \sem{Q}_G \big(\sem{C}_G(T)\big)$\\
        where~$C$ is a clause and~$Q$ is a query.
\end{itemize}

\subsection{Semantics of clauses}
The meaning of Cypher clauses is again functions that take tables to
tables.
Matching clauses are essentially
pattern matching statements: they are of the form $\kwcode{OPTIONAL}\
\kwcode{MATCH}\ \pathpatternlist\ \kwcode{WHERE}\ \expr$. Both \kwcode{OPTIONAL}
and \kwcode{WHERE} could be omitted. The key to their semantics is pattern
matching, in particular $\match(\bar \pi, G, u)$ described in
Section~\ref{sec:pattern} (see Equation~(\ref{match-eq}), page \pageref{match-eq}).

The \kwcode{MATCH} clause extends the set of field names of $T$ by adding to it
field names that correspond to names occurring in the pattern but not in $u$. It
also adds tuples to $T$, based on matches of the pattern that are found in
graphs. \kwcode{UNWIND} is another clause that expands the set fields, and
\kwcode{WITH} clauses can change the set of fields to any desired one.
The \kwcode{WHERE} subclause also defines a table-to-tables function that
filters lines according to the evaluation of an expression; it is not a proper
clause because of its interaction with \kwcode{OPTIONAL\ MATCH} clauses.

\paragraph{Matching clause} The semantics of \sqlkw{MATCH} clauses is defined below; the
semantics of \sqlkw{WHERE} subclause is defined afterwards.

\begin{itemize}[itemsep=2mm]
\item $\displaystyle \sem{\kwcode{MATCH}\ \bar \pi}_{G}(T)  = \biguplus_{u\in T}
\{u \cdot u' \mid u' \in \textsf{match}(\bar \pi, G, u)\}$

\item $\sem{\kwcode{MATCH}\ \bar \pi \ \kwcode{WHERE}\ e}_{G}(T)
  = \sem{\kwcode{WHERE}\ e} \Big(\sem{\kwcode{MATCH}\ \bar \pi}_{G}(T) \Big)$

\item $\displaystyle \sem{\kwcode{OPTIONAL MATCH}\ \bar \pi
	\ \kwcode{WHERE}\ e}_{G}(T) $\\\hspace*{0cm plus 1fill} $\displaystyle = \biguplus_{u \in T}
					\left \{
						\begin{array}{ll}
							\sem{\kwcode{MATCH}\ \bar \pi\  \kwcode{WHERE}\ e}_G(\set{u}) & \text{ if }  \sem{\kwcode{MATCH}\ \bar \pi\ \kwcode{WHERE}\ e}_G(\set{u}) \neq \emptyset \\
							(u,(\text{free}(u,\bar \pi):\nv)) & \text{ otherwise} \\
						\end{array}
					\right.$

\item $\displaystyle \sem{\kwcode{OPTIONAL MATCH}\ \bar \pi}_{G}(T)  = \sem{\kwcode{OPTIONAL MATCH}\ \bar \pi\ \kwcode{WHERE}\ \true}_{G}(T)$
\end{itemize}

\begin{example}
  Let $G$ be the property graph defined in Figure \ref{fig:teachers}. consider
  the clause \kwcode{MATCH} $\pi$, where $\pi$ is the pattern
  \begin{center}
    \lstinline[style=cypher]!(x) -[:KNOWS*]-> (y)!
  \end{center}
  Let $T$ be the table $\set{(x:n_1);(x:n_3)}$ with a single field $x$. We show
  how to compute $\sem{\kwcode{MATCH}\ \pi}_G(T)$.

  Note that $\rigid(\pi)$ is the (infinite) set of all rigid paths
  $\pi_m=(\ltor,\nil,\{\textsc{knows}\},m,m)$, for $m > 0$.
  These can only be satisfied by paths with exactly $m$ distinct
  relationships. Since $G$ only contains $3$ relationships, only
  $\pi_1$, $\pi_2$ and $\pi_3$ can contribute to the   result.

  Let $u = (x:n_1)$, $\pi' = \pi_1$ and $p =
  n_1r_1n_2$. Then $\freevars(\pi_1)-\dom(u) = \set{y}$, and thus $u'$ must
  be a record over the field $y$. One can easily check that
  $(n_1r_1n_2,G,(x:n_1,y:n_2))\models\pi_1$. In fact $n_2$
  is the only suitable value for $y$, and thus the contribution of this
  specific triple $u,\pi',p$ to the final result is precisely $\set{(x:n_1 ,
    y:n_2)}$.

  No path $p$ other than $n_1r_1n_2$ can contribute a record
  in the case where $u = (x:n_1)$ and $\pi' = \pi_1$. Indeed, $\pi_1$ requires
  $p$ to be of length $1$,  and start at $x$, which $u$
  evaluates to be $n_1$.  By a similar reasoning,
  we can compute the  contribution of the following triples:
  \begin{itemize}
  \item $(x:n_1,y:n_3),\,\pi_2,\,n_1r_1n_2r_2n_3$ { }yields{ } $(x:n_1,y:n_3)$;
  \item $(x:n_1,y:n_4),\,\pi_3,\,n_1r_1n_2r_2n_3r_3n_4$ { }yields{ } $(x:n_1,y:n_4)$;
  \item $(x:n_3,y:n_4),\,\pi_1,\,n_3r_3n_4$ { }yields{ } $(x:n_3,y:n_4)$;
  \end{itemize}
  and show that the contributions of all other possible combinations of records,
  paths and patterns are empty. This tells us that \linebreak $\sem{\kwcode{MATCH}\
    \pi}_G(T)$ is the following table:
  \begin{center}
    \begin{tabular}{cc}
      \hline
      $x$ & $y$ \\
      \midrule
      $n_1$ & $n_2$ \\
      $n_1$ & $n_3$ \\
      $n_1$ & $n_4$ \\
      $n_3$ & $n_4$ \\
      \hline
    \end{tabular}
  \end{center}
\end{example}

\paragraph{Where subclause}
  Although \sqlkw{WHERE} is not a clause per say, its semantics is also a table
  to table function.

\begin{itemize}
  \item $\sem{\kwcode{WHERE}\ e}_G(T)  = \set{u \in T \ |
  \ \sem{e}_{G,u} = \true}$
\end{itemize}

\paragraph{With clause}

Similarly to the description of the semantics of \sqlkw{RETURN} queries, we make
the assumption that the fields of $T$ are
$b_1,\ldots,b_q$. Our convention about the names $a_i'$ are exactly the same as
for queries (see above), except that $a_i'=\alpha(e_i)$ only if $e_i$ is a name.

\begin{itemize}[itemsep=2mm]
  \item $\sem{\kwcode{WITH}\ *}_G(T)  =  T  \\\text{ if }T\text{ has a least one field}$

  \item $\sem{\kwcode{WITH}\ e_1 \ [\kwcode{AS}\ a_1], \ldots, e_m
  \ [\kwcode{AS}\ a_m]}_G(T)  $\\\hspace*{0cm plus 1fill}$\displaystyle{}=
\biguplus_{u\in T} \set{(a_1' : \sem{e_1}_{G,u}, \ldots, a_m' : \sem{e_m}_{G,u})}$

  \item $\sem{\kwcode{WITH}\ *, e_1 \ [\kwcode{AS}\ a_1], \ldots, e_m
  \ [\kwcode{AS}\ a_m]}_G(T)  $\\\hspace*{0cm plus 1fill}$=
\sem{\kwcode{WITH}\ b_1 \ \kwcode{AS} \ b_1,\ldots, b_q \ \kwcode{AS}\ b_q,
  e_1 \ [\kwcode{AS}\ a_1], \ldots, e_m \ [\kwcode{AS}\ a_m]}_G(T)$

  \item $\sem{\kwcode{WITH}\ \retarg\ \kwcode{WHERE}\ e}_{G}(T)
              =
\sem{\kwcode{WHERE}\ e}_{G}\Big(\sem{\kwcode{WITH}\ \retarg}_{G}(T)\Big)$

\end{itemize}

\paragraph{Unwind Clause}

\begin{itemize}[itemsep=2mm]
  \item $\displaystyle\sem{\kwcode{UNWIND} \ e \ \kwcode{AS} \ a}_G(T) = \biguplus_{u\in
  T}\biguplus_{v\in E_u}\set{(u,a:v)}~, $\\[1mm]\hspace*{0mm plus 1fill} $\text{where}\ \
E_u  = \left\{\!\! \begin{array}{ll}
					\biguplus_{0 \leq i < m} \set{v_i} & \text{if } \sem{e}_{G,u} = \text{list}(v_0,\ldots,v_{m-1}) \\
					\set{} & \text{if } \sem{e}_{G,u} = \text{list}() \\
					\set{\sem{e}_{G,u}} & \text{otherwise }
				\end{array}\right.$
\end{itemize}

\bibliography{biblio}

\end{document}